\documentstyle{elsart}
\input psfig

\begin{document}
\begin{frontmatter}

\title{Magnetic-Field-Driven Superconductor-Insulator-Type Transition
in Graphite}

\author[leipzig]{H. Kempa},
\author[leipzig]{Y. Kopelevich\thanksref{yako}}, 
\author[leipzig]{F. Mrowka},
\author[leipzig]{A. Setzer},
\author[leipzig]{J. H. S. Torres\thanksref{yako}}, 
\author[leipzig]{R. H\"ohne} and 
\author[leipzig]{P. Esquinazi} 
\address[leipzig]{Department of Superconductivity
and Magnetism,  Institut  f\"ur Experimentelle Physik II, 
Universit\"at Leipzig, Linn\'estr. 5, D-04103  Leipzig, Germany}
\thanks[yako]{On leave from Instituto de Fisica, Unicamp,
13083-970 Campinas, Sao Paulo, Brasil.}

\begin{abstract}
A magnetic-field-driven transition from metallic- to
semiconducting-type  behavior in the basal-plane resistance takes
place in highly oriented pyrolytic graphite  at a field $H_c \sim
1~$kOe applied along the hexagonal c-axis. The analysis of the data  
reveals a striking similarity between this transition and that
measured in thin-film superconductors  and Si MOSFET's. However, in
contrast to those materials, the transition in  graphite is observable at
almost two orders of magnitude higher temperatures. 
\end{abstract}

\end{frontmatter}


The anomalous carrier density, temperature, magnetic  and electric
field dependence of the resistivity observed in various
two-dimensional (2D) electron and hole systems \cite{sim} as well as
in superconducting thin films
\cite{yaz,mar} remains still without an accepted explanation. 
Transport measurements in Si MOSFET's with large enough carrier
density  reveal that the resistivity drops by a large  factor as the
temperature decreases below a temperature $T \sim \frac{1}{3}T_F$
($T_F$  is the Fermi temperature) indicative of a clear metallic
behavior. Taking into account the observed  magnetic field
suppression of the metallic behavior \cite{sim}, the possible
existence of a superconducting phase in two dimensions in Si MOSFET's
has been suggested \cite{phi} based mainly on the field-induced
superconductor-insulator (SI) transition scaling analysis \cite{fis}.

Recent  studies of highly oriented pyrolitic graphite (HOPG) suggest
the occurrence of high-temperature superconducting correlations in
this material \cite{kope1,kope2}.  Noting that graphite is a quasi-2D
semimetal with a low carrier  density (electrons and holes) so that
Coulomb interaction effects can be important (the concentration of
the majority carriers is $\sim  2 \times 10^{18}~$cm$^{-3}$ and of
the minority carriers is $\sim 6 \times 10^{16}$~cm$^{-3}$
\cite{dre2,kel,wil,sha}) it seems  reasonable to ask   whether some
of the ideas regarding   superconductivity in MOSFET's
\cite{sim,phi,zha,rice,bel} may be also applicable to graphite.  

In this work we have studied both   magnetoresistance and
magnetization of HOPG samples with the  basal plane resistivity
$\rho$  ranging from $\rho(T = 300~$K)$ = 5~\mu\Omega$cm to $135 \mu\Omega$cm.  
Our results demonstrate that a field $H =
H_c \sim 1~$kOe applied parallel to the hexagonal  c-axis, induces a
transition from metallic- (d$R/{\rm d}T > 0, R$ is the measured
resistance)  to semiconducting-type (d$R/{\rm d}T < 0$)  behavior.
The analysis of the transport data, in particular the obtained
scaling reveals a striking similarity to  that  measured
in 2D superconductors \cite{yaz,mar,heb,zan,mas} and Si MOSFET's
\cite{sim,phi}.   Here we concentrate  mainly on the data obtained
for the sample with the lowest resistivity (sample HOPG-3).

HOPG samples were synthesized at temperatures $T \sim 2700 \ldots 
3000~^{\rm o}$C  and pressure $P  \sim 10\ldots 30~$MPa at the 
Research Institute ``Graphite'' (Moscow) and characterized by means 
of x-ray diffraction (the analysis of the rocking curves FWHM
indicate for the different samples: HOPG-1: $1.4^o$, HOPG-2: $1.2^o$,
HOPG-3: $0.5^o$), SEM, STM, and spectrographic analysis techniques 
\cite{kope1,kope2}. Dc magnetization $M(H,T)$ measurements were
performed with the SQUID magnetometer  MPMS7 (Quantum Design). The dc
resistance $R(H,T)$ was measured in the temperature  interval 2~K - 300~K
 in applied magnetic field $0 \le H \le 9~$T using a standard 4-probe method 
with both current and voltage leads patterned on the 
sample surface. Resistivity values were obtained with a lead geometry
that assures an uniformly distributed current over  the sample cross
section. The amplitude of the applied current was between 10~$\mu$A 
and  1~mA at which no Joule heating was detected. In all transport
measurements the current was inverted in  order to eliminate 
thermoelectric effects.  

\begin{figure}
\centerline{\psfig{file=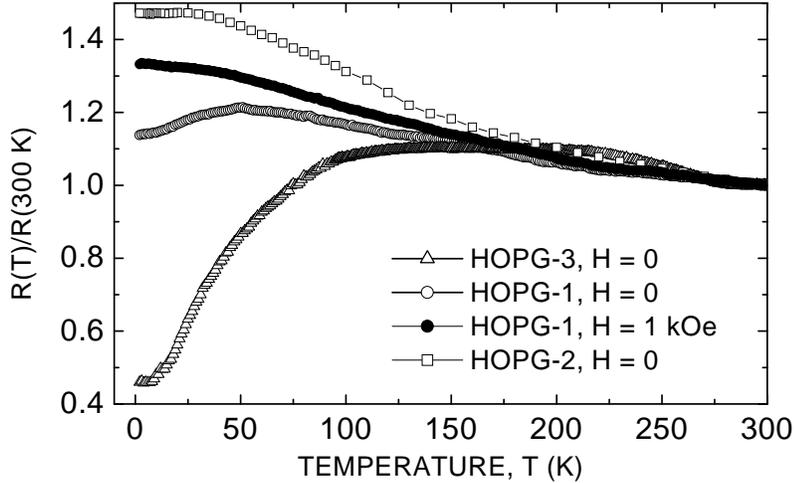,height=3.2in}}
\caption{Temperature dependence of the reduced basal-plane 
resistance $R(T)$ obtained for  three graphite samples (HOPG-1, -2, and -3) 
in zero applied magnetic field and  also in $H = 1~$kOe for HOPG-1 sample.} 
\label{resvst} \end{figure}
Figure~\ref{resvst}~shows~the normalized resistance $R(T)/R(300~$K)
measured for samples with  $\rho(300~$K)$ = 45 \mu\Omega$cm  (sample
HOPG-1), $\rho(300$~K) = 135~$\mu\Omega$cm (sample HOPG-2),  and 
$\rho(300$~K) = 5 $\mu\Omega$cm (sample HOPG-3). As can be seen from
Fig.~\ref{resvst}, at $T > 200~$K  all three  samples show a
semiconducting-type temperature dependence. At lower temperatures
$R(T)$ obtained for the less resistive  samples (HOPG-3  and HOPG-1) crosses
over to a metallic behavior (d$R/{\rm d}T > 0)$  below $T \sim 200~$K
and $T \sim 50~$K, respectively.  However, in the sample HOPG-2 with
the highest resistivity, d$R/{\rm d}T < 0$  down to $T \sim 30~$K
below which $R(T)$ saturates. We stress that the $R(T)$-curves  shown
in Fig. \ref{resvst} are not specific to our samples: both metallic-
and nonmetallic-type $R(T)$  generally occur in HOPG depending on the
heat treatment \cite{dre2,kel}. Figure \ref{resvst} also  
illustrates that a weak ($\sim 1~$kOe) magnetic field $H ||$c-axis
suppresses the metallic-type state in the  HOPG-1 sample.

\begin{figure}
\centerline{\psfig{file=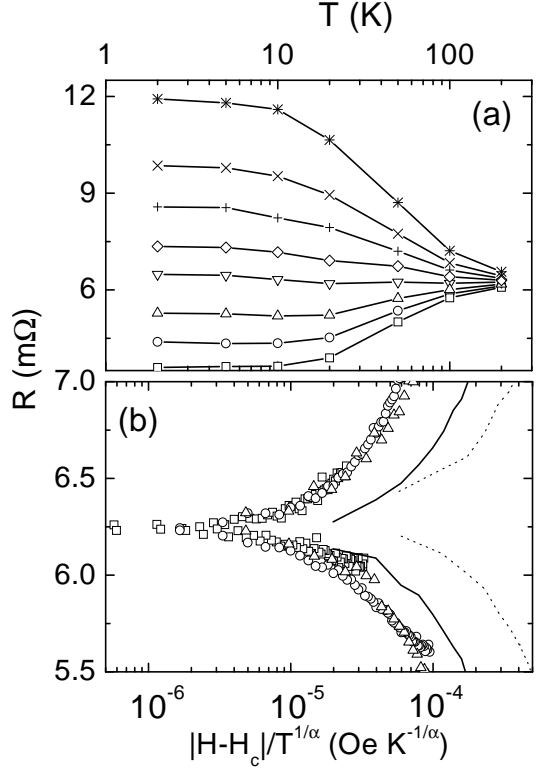,height=4.4in}}
\caption{(a) $R(T)$ obtained for HOPG-3 for selected magnetic fields
using magnetoresistance data: $(\Box) H = 500~$Oe, $(\circ) 700~$Oe, 
$(\triangle)  900~$Oe, $(\bigtriangledown) 1140~$Oe, $(\Diamond)
1300~$Oe, $(+) 1500~$Oe, $ (\times) 1700~$Oe, $(\ast) 2000~$Oe. (b)
Resistance measured at  $T = 200 {\rm K} (\Box),  100 {\rm K}
(\circ), 50 {\rm K} (\triangle), 20 {\rm K} (-),$ and  10K (- -) as a
function  of the scaling variable where $H_c = 1140~$Oe and  $\alpha
= 0.65 \pm 0.05$.} \label{ressca} \end{figure}

\begin{figure}
\centerline{\psfig{file=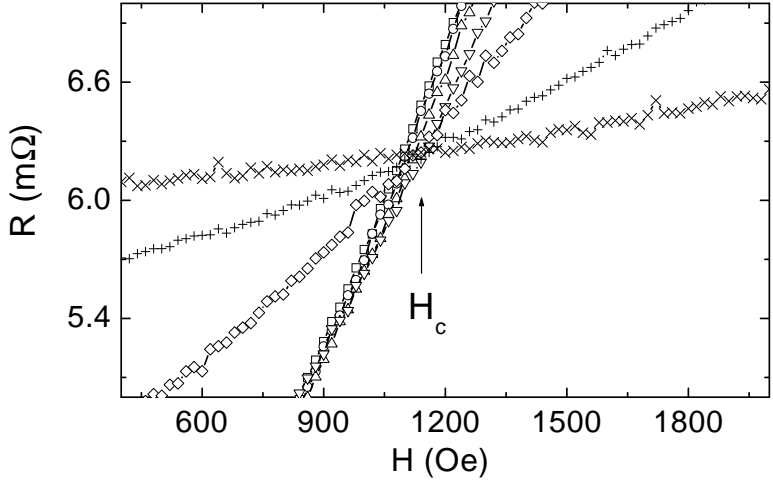,height=3.2in}}
\caption{Magnetoresistance $R(H)$ isotherms measured in HOPG-3:
$(\Box)~T=2~$K,  $(\circ) 5~$K,  $(\triangle) 10~$K,
$(\bigtriangledown) 20~$K, $(\Diamond) 50~$K, $(+)  100~$K, $(\times)
200~$K. $H_c = 1140~$Oe is the ``crossover" field.} \label{cross}
\end{figure}

A pronounced magnetic-field-induced transition from metallic- to
semiconducting-type $R(T)$ is observed
 in sample HOPG-3, as
shown in Fig.~\ref{ressca}(a). The $R(T)$ data plotted in
Fig.~\ref{ressca}(a) for various applied magnetic fields  were
obtained from isothermal  magnetoresistance $R(H)$ measurements.
Figure~\ref{cross} shows  several $R(H)$ isotherms in the  vicinity
of the ``crossover''  field $H_c = 1140~$Oe at which the $R(H)$ data
obtained at temperatures 50~K, 100~K and 200~K intersect. 

Noting that the results of Fig.~\ref{ressca}(a) resemble very much
the resistance behavior in the vicinity of the
magnetic-field-driven metal-insulator (MI)  
 \cite{sim,phi} and  SI \cite{yaz,mar,heb,zan,mas}
transitions   we apply here the same scaling approach
\cite{yaz,mar,phi,fis,heb,mas} to the transition measured in HOPG.  According to
the scaling analysis, the resistance in the critical regime of the
quantum SI transition is given by $R(\delta, T) = R_c
f(|\delta|/T^{1/z\nu})$, where $R_c$ is the resistance at  the
transition,  $f(|\delta|/T^{1/z\nu})$ the scaling function such that
$f(0) = 1$,  $z$ and $\nu$ are critical exponents, and $\delta$ is
the  deviation of a variable parameter from its critical value. With
$\delta = H - H_c$, we have plotted  in  Fig.~\ref{ressca}(b) $R$ vs.
$|\delta|/T^{1/\alpha}$,  where $\alpha = 0.65 \pm 0.05$ was obtained
from the log-log plot of  (d$R/{\rm d}H)|_{H_c}$ vs. $T^{-1}$.  The
collapse of the resistance data into two distinct branches, below  
 and above the critical field $H_c$, is obtained  in
the temperature range 50~K - 200~K, see Fig.~\ref{ressca}(b). At  $T
< 20~$K, where the resistance saturation is apparent, see
Fig.~\ref{ressca}(a), a clear deviation  from the scaling is evident, 
reminiscent of the behavior observed in  amorphous Mo-Ge films
\cite{mas} where still unknown dissipation effects \cite{mas,shi} 
lead to the saturation of the resistance at low temperatures.  Interestingly,
 the obtained value of the exponent $\alpha = 0.65 \pm 0.05$  for
HOPG  coincides with that found in the scaling analysis of both the
magnetic-field-tuned MI-type  transition in Si MOSFETS's ($\alpha =
0.6 \pm 0.1$ \cite{phi})  and the SI transition in ultrathin a-Bi films
($\alpha =  0.7 \pm 0.2$ \cite{mar}).

We would like to  stress the similarities in the transport properties
of HOPG and those of a 2D electron gas as in Si MOSFET's: (1) The
occurrence of metallic or semiconducting behavior in graphite at $H = 0$ (see
Fig.~\ref{resvst}) is very sensitive to the carrier density which can
be changed by, e.g., annealing 
\cite{kel}. (2) The characteristic temperature $T^*$ ($\sim 100$~K for
sample HOPG-3) below which $\rho(T)$ strongly decreases is a
considerable fraction of the Fermi temperature $T_F \sim 250~$K \cite{sha}. 
(3) There is a clear suppression of the
conducting phase by a magnetic field. 
(4) There exists a critical scaling with a remarkably  similar exponent
$\alpha$.
  
There is, however, an important
difference between the results in HOPG and those obtained in Si
MOSFET's. Whereas the magnetoresistance in Si MOSFET's is 
independent of the direction of the applied magnetic field \cite{sim},
$\Delta R/R$  measured in HOPG is two  orders of magnitude larger for
$H ||~$c-axis as compared to the perpendicular orientation. In  other
words, while a spin-polarization mechanism seems responsible for the
transition  in Si MOSFET's, orbital effects are dominant and
influence the transition in  HOPG. 

\begin{figure}
\centerline{\psfig{file=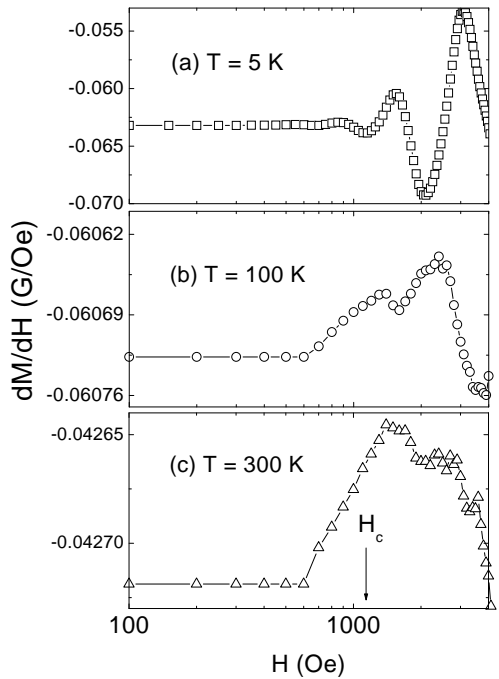,height=4.0in}}
\caption{Field derivative of the measured magnetization (without
any subtraction) as a function
of applied field for sample HOPG-3 at three temperatures.}
\label{dhvh}
\end{figure}
We note that the scaling shown in Fig.~\ref{ressca}(b) can be
related to the field-tuned quantum transition if the  critical regime
extends to $T \sim 200~$K, at least. Although this may appear unlikely,
the special characteristic of graphite must be taken into account:
Because of the low density and extremely small effective
masses of the carriers in graphite, as  well as its quasi two-dimensional
nature, one expects a strong enhancement of quantum  effects. 
Figure \ref{dhvh} shows the experimental evidence for quantum
oscillations due to de Haas-van Alphen (DHVA) effect at
$T = 5~$K (a), 100~K (b) and 300~K (c). The occurrence of
DHVA oscillations at low fields arising from the minority
carriers was already reported in the literature \cite{dre2,kel,wil}.
It is interesting to note, however, that these oscillations become
particularly clear at $H \ge H_c \sim 1~$kOe where the SI-type transition
takes place, see Fig.~\ref{ressca}.

The following points should be considered  regarding the origin of
possible superconducting correlations in HOPG. (a) We would like to
emphasize the potential importance of the minority carriers in HOPG.
The effective mass  of the minority carriers $m^* \sim 0.004 m_0$ is
10 to 15 times smaller than the effective mass of the majority  
holes  ($m^* \sim 0.04 m_0$) and electrons ($m^* \sim 0.06 m_0$)
\cite{dre2,kel,wil,sha} ($m_0$ is the  free-electron mass). This may
lead to  superconducting instabilities in the electron-hole liquid as
discussed in Refs.\cite{vig,ric,abr}.  (b) On  the other hand, the
coexistence and interplay of superconducting and 
ferromagnetic states \cite{kope1,exp} may indicate
the relevance of p-wave pairing  mediated by ferromagnetic spin
fluctuations \cite{bel,mur}.

To conclude, the present work provides evidence that at least some of
the  physical concepts proposed to describe the magnetic-field-tuned
SI-type transition in various 2D
systems should also be applicable to graphite. The  ``reentrant''
metallic-type behavior observed in the regime of lowest Landau level
quantization for  both HOPG \cite{kope2} and Si
MOSFET's \cite{kra} further suggests a deep similarity of the
physical  processes operating in these systems. 

\ack
We are grateful to 
A. A. Abrikosov, A. Gerber, I. Ya.  Korenblit, I. Luk'yanchuk, A. C.
Mota,  S. Moehlecke, K. A. M\"uller, A. Shelankov,  M. Sigrist, Z. Tesanovic, and
M. Ziese for informative discussions. The authors thank A. S.
Kotosonov for providing the samples and V. V. Lemanov for
collaboration. This work is supported by the Deutsche 
Forschungsgemeinschaft under DFG IK 24/B1-1 (project H),  and was
partially  supported by the DAAD and CAPES Proc. No. 077/99 and CNPq proc.
No. 301216/93-2. F.M. is
supported by the German-Israeli Foundation under G-553-191.14/97.

\end{document}